\documentclass[a4paper]{jpconf}
\pdfoutput=1
\usepackage{graphicx,amssymb}
\graphicspath{{figs/}}

\begin{document}
\title{The current status of orbital experiments for UHECR studies}

\author{M~I~Panasyuk$^{1,2}$, 
	M~Casolino$^{3,4}$, 
	G~K~Garipov$^1$,
	T~Ebisuzaki$^{3}$,
	P~Gorodetzky$^{5}$,
	B~A~Khrenov$^1$, P~A~Klimov$^1$, 
	V~S~Morozenko$^1$,
	N~Sakaki$^{6}$, 
	O~A~Saprykin$^7$, S~A~Sharakin$^1$, 
	Y~Takizawa$^{3}$, 
	L~G~Tkachev$^8$, I~V~Yashin$^1$ and
	M~Yu~Zotov$^1$}

\address{$^1$ D.V. Skobeltsyn Institute of Nuclear Physics, M.V. Lomonosov
	Moscow State University (SINP MSU), Moscow 119991, Russia}
\address{$^2$ Faculty of Physics, M.V. Lomonosov Moscow State University,
	Moscow 119991, Russia}
\address{$^3$ RIKEN, 2-1 Hirosawa, Wako, 351-0198 Japan}
\address{$^4$ Istituto Nazionale di Fisica Nucleare~-- Sezione di Roma Tor
	Vergata, Italy}
\address{$^5$ APC, Univ Paris Diderot, CNRS/IN2P3, CEA/Irfu, Obs de Paris,
	Sorbonne Paris Cit\'e, France}
\address{$^6$ Karlsruhe Institute of Technology (KIT), 76021 Karlsruhe, Germany.
	Now at: Osaka City University, Japan}
\address{$^7$ Central Research Institute of Machine Building (TsNIIMash),
	Korolev 141070, Russia}
\address{$^8$ Joint Institute for Nuclear Research (JINR), Dubna 141980, Russia}

\ead{panasyuk@sinp.msu.ru}

\begin{abstract}

Two types of orbital detectors of extreme energy cosmic rays are being
developed nowadays: (i)~TUS and KLYPVE with reflecting optical systems (mirrors)
and (ii)~JEM-EUSO with high-transmittance Fresnel lenses. 
They will cover much
larger areas than existing ground-based arrays and almost uniformly monitor the
celestial sphere. The TUS detector is the pioneering mission developed in SINP
MSU in cooperation with several Russian and foreign institutions. It has
relatively small field of view ($\pm4.5^\circ$), which corresponds to a ground
area of $6.4\cdot10^3$~km$^2$. The telescope consists of a Fresnel-type
mirror-concentrator ($\sim2$~m$^2$) and a photo receiver (a matrix of
$16\times16$ photomultiplier tubes). It is to be deployed on the Lomonosov
satellite, and is currently at the final stage of preflight tests. Recently, SINP
MSU began the KLYPVE project to be installed on board of the Russian segment of
the ISS.  The optical system of this detector contains a larger primary mirror
(10~m$^2$), which allows decreasing the energy threshold. The total effective
field of view will be at least $\pm14^\circ$ to exceed the annual exposure of
the existing ground-based experiments. Several configurations of the detector
are being currently considered. Finally, JEM-EUSO is a wide field of view
($\pm30^\circ$) detector.  The optics is composed of two curved double-sided
Fresnel lenses with 2.65~m external diameter, a precision diffractive middle lens and a
pupil. The ultraviolet photons are focused onto the focal surface, which consists of
nearly 5000 multi-anode photomultipliers.  It is developed by a large
international collaboration. All three orbital detectors have multi-purpose
character due to continuous monitoring of various atmospheric phenomena. The
present status of development of the TUS and KLYPVE missions is
reported, and a brief comparison of the projects with JEM-EUSO is given.

\end{abstract}

\section{Introduction}

The nature and origin of extreme energy cosmic rays (EECRs, those with energies
$\gtrsim50$~EeV\footnote{$1~\mathrm{EeV}=10^{18}~\mathrm{eV}.$}) remains one of
the greatest puzzles of modern astrophysics after more than 50 years since their
first registration~\cite{Linsley-1e20-1963}.  In this introduction, we will
briefly review a number of important experimental results on the subject that
were obtained in recent years.  These include results on the energy spectrum,
mass composition and anisotropy of ultra-high energy cosmic rays (UHECRs)
obtained with the Pierre Auger Observatory (Auger for brevity), located in
Argentina (South hemisphere); High Resolution Fly's Eye (HiRes) and the
Telescope Array Project (TA), both located in Utah, USA (North hemisphere).
Then we recall the main idea behind observation of EECRs from space and
review three orbital projects aimed to become the next generation of EECR
experiments that are currently under active development mostly focusing on TUS
and KLYPVE, but also briefly discussing the JEM-EUSO mission.
We aim to explain how these projects are able to open new
possibilities to solve the EECR puzzle.

The most important result relates to the all-particle energy
spectrum of UHECRs.  It was found by HiRes~\cite{Hires-GZK-2007,HiRes-GZK-2008},
Auger~\cite{Auger-GZK} and confirmed by TA~\cite{TA-GZK-2013} that the flux of
cosmic rays is strongly suppressed at energies $\gtrsim50$~EeV.  A joint working
group of the Pierre Auger, Telescope Array and Yakutsk Collaborations concluded
that the energy spectra determined by Auger and TA are consistent in
normalization and shape within systematic uncertainties after energy scaling
factors are applied~\cite{WG-spectrum-2013}: both spectra demonstrate an
``ankle'' at around 5~EeV and a cut-off at $\gtrsim50$~EeV.  It was found though
that the cut-off takes place at lower energies in the spectrum by Auger and is
steeper than the TA one.

The cut-off was found at the energy scale predicted by
Greisen~\cite{Greisen-1966} and independently by Zatsepin and
Kuz'min~\cite{ZK-1966}, who demonstrated that the flux of protons will be
strongly suppressed at energies $\gtrsim50$~EeV due to the photopion production
on the cosmic microwave background radiation.  Still, the discrepancy in
the energy spectra obtained by Auger and TA does not allow concluding if the
observed cut-off is due to the GZK effect, due to the maximum energy attained in
accelerators of EECRs, or a combination of both, see~\cite{Harari-GZK-2014} for
an in-depth discussion.  It does not either allow one to make a definite
conclusion concerning the mass composition of the primary particles: are they
protons or heavier nuclei, e.g.~iron.

Another discrepancy between results of the Auger and TA experiments further
complicates the situation with the composition of UHECRs.  Namely, the Auger
data witness in favour of the mass composition becoming heavier at energies
above 3~EeV~\cite{Auger-mass-2013} while the TA data are consistent with a
light, dominantly protonic composition~\cite{TA-mass-2013,TA-mass-2015} but
incompatible with a pure iron composition.  The TA results are also in agreement
with previous HiRes stereo measurements.  The elongation rate and mean values of
the depth of the shower maximum measured at TA are in good agreement with Auger
data~\cite{TA-mass-2015}.  Though one should keep in mind that, on the one hand,
the measurement of masses of primary particles is one of the most difficult
tasks in the physics of UHECRs since it is based on hadronic interaction models
at energies far beyond the reach of man-made accelerators like the LHC. On the
other hand, Auger and TA employ different analysis techniques and different cuts
selecting data for the analysis, so that a direct comparison of the results of
the two experiments can be misleading.  A work on a detailed comparison of mass
composition obtained with Auger and TA is in progress~\cite{Auger-TA-2013}.

One more aspect of cosmic ray physics that attracts great attention is
anisotropy of arrival directions of EECRs. It is one of the key elements in a
search for possible sources but also useful for examining mass composition of
EECRs independently of measurements of the shower depth.  The most intriguing
new result in the field is the discovery of a ``hotspot'' of cosmic rays
with energies above 57~EeV made by the TA experiment in the Northern
hemisphere~\cite{TA-hotspot-2014}.  The hotspot is a cluster of 19 (of 72
registered) events within a $20^\circ$ radius circle centered at (equatorial
coordinates) $(\alpha,\delta)=(146.\!^\circ7,43.\!^\circ2)$ with an expected
number of events equal to 4.49, which gives a (pre-trial) statistical
significance of~$5.1\sigma$. The (post-trial) statistical significance of such a
hotspot appearing by chance was estimated to be~$3.4\sigma$.  The hotspot is
located near the Supergalactic plane but there are no known specific sources
behind it.  It was suggested by the TA Collaboration that the hotspot may be
associated with the closest galaxy groups and/or the galaxy filament connecting
the Milky Way with the Virgo cluster; or, if EECRs are heavy nuclei in
accordance with the Auger results, the hotspot events may originate in the
Supergalactic plane and be deflected by the magnetic fields.  Recently, it has
been argued that the nearby starburst galaxy M82 and the bright nearby blazar
Mrk~180 (or one of them) are likely sources of the TA hotspot assuming its
events have a pure composition~\cite{Kusenko_etal-hotspot-2014}.  Still, the
available statistics, the ambiguity in the mass measurements of EECRs and the
insufficient knowledge of extragalactic magnetic fields do not allow making a
definite conclusion.

The Pierre Auger Collaboration is also performing intensive studies of
anisotropy.  In particular, a correlation is reported between the arrival
directions of cosmic rays with energies above 55~EeV and the distribution of
active galactic nuclei (AGN) within 75~Mpc, among them the Centaurus~A (Cen~A)
radiogalaxy, located at less than 4~Mpc
distance~\cite{Auger-science-2007,Auger-ApP-2008,Auger-ApP-2010}.  More
recently, Auger presented results of a whole number of tests aimed to search for
signals of anisotropies of cosmic rays with energies above 40~EeV (with 231 of
them having energies $\ge52$~EeV and zenith angles
$<80^\circ$)~\cite{Auger-aniso-2014}.  None of the tests performed revealed a
statistically significant deviation from isotropy but it was found that there is
a certain excess of events with energies $\ge58$~EeV around the direction to
Cen~A and around AGN of the Swift catalogue within 130~Mpc and brighter than
$10^{44}$~erg/s. In both cases, the probability of arising the excess by chance
was estimated to be of the order of 1.3--1.4\%.  No hotspot similar to the one
observed by TA was found.  On the other hand, a similar study performed by TA
using EECRs with energies $\ge40$~EeV registered by the surface detector during
the first 40 months of its operation did not reveal any statistically
significant correlations with AGN from a number of survey
catalogues~\cite{TA-exgal-2013}.

Another direction of anisotropy studies is a search for large-scale anisotropy
of EECRs. Recently, Auger presented results of two Rayleigh analyses, one in the
right ascension and one in the azimuth angle distributions, of the arrival
directions of events with energy above 4~EeV with zenith angle up to
$80^\circ$~\cite{Auger-largescale-2014}.  The largest departure from isotropy
was found in the $E > 8$~EeV energy bin, with an amplitude for the first
harmonic in right ascension $(4.4\pm1.0)\times10^{-2}$, which has a chance
probability of $6.4\times10^{-5}$.  It was found that assuming the only significant
contribution to large-scale anisotropy is from the dipolar component, the
observations above 8~EeV correspond to a dipole of amplitude $0.073\pm0.015$
pointing to $(\alpha,\delta)=(95^\circ\pm13^\circ, -39^\circ\pm13^\circ)$, thus
supporting an earlier result obtained for events with zenith angles
$<60^\circ$~\cite{Auger-deAlmeida-2013}.

Telescope Array has also performed an analysis of large-scale anisotropy of
cosmic rays at the highest end of the spectrum, though a totally different approach was
employed~\cite{TA-large-2013}.  Assuming pure proton composition, it was found
that the distribution of events with energies $>10$~EeV and $>40$~EeV is compatible
with the hypothesis of isotropy and incompatible with the distribution of local
large-scale structure (LSS) at smearing angles smaller than $\sim20^\circ$ and
$\sim10^\circ$ respectively.  To the contrary, the data set composed of events
with energy above 57~EeV was found to be compatible with the LSS model and
incompatible with isotropy at (pre-trial) statistical significance
$\sim3\sigma$.

Anisotropy studies present a complicated task because none of the experiments
observes the whole celestial sphere but a proper determination of the full set
of multipole coefficients requires full-sky coverage.  That is why a joint
analysis of large-scale anisotropy of UHECRs with energies above 10~EeV
performed by the TA and Pierre Auger collaborations is of special
interest~\cite{Auger-TA-largescale-2014}.  A declination band commonly covered
by both experiments ($-15^\circ<\delta<25^\circ$) was used for cross-calibrating
the fluxes.  The multipolar expansion of the UHECR flux obtained in the study
allowed performing a series of anisotropy searches but no significant deviation
from isotropy was revealed at any angular scale.  It was later found that the
obtained angular power spectrum contradicts the hypothesis that the distribution
of UHECRs at energies above 10~EeV follows the LSS model if a purely protonic
primary composition is assumed, in that amplitudes of low multipoles in the data
are significantly lower than those calculated from the LSS
model~\cite{Tinyakov-Urban-2014}.  One of the possibilities to explain the
result is that the primaries are heavy, as suggested by the Auger collaboration
results.  

Nowadays, obtaining definite conclusions concerning anisotropy of EECRs becomes
one of the most important tasks.  As is clear from the above, a considerable
part of the difficulties in finding the sources of EECRs is due to (i)~the
limited statistics of events and (ii)~the incomplete and non-uniform coverage of
the celestial sphere by any of the modern experiments.  Thus one needs to
increase the statistics of EECRs providing a uniform exposition of the sky.
This is exactly the field where orbital experiments open new horizons of
research (not to mention a whole number of other scientific tasks related 
to astrophysics, fundamental physics and atmosphere sciences, which we do not
discuss here) being complementary to ground-based detectors.

\section{Observation of EECRs from space}

The atmosphere of the Earth acts as a huge calorimeter for detecting UHECRs due
to extensive air showers (EAS) produced by primary particles moving nearly at
the speed of light.  The secondaries of an EAS cascade ionize and excite
molecules of nitrogen in the air, which then radiate photons in the range
$\approx330$--400~nm.  This near-ultraviolet (UV) air fluorescence light is
produced isotropically and can be used to record the development of an EAS.  The
number of photons emitted at any point along the EAS track is proportional to
the number of charged particles in the cascade, mostly electrons and positrons.
Additional information can be obtained from the Cherenkov light diffusively
reflected at the surface of the Earth.  Both can be used to reconstruct the
energy and arrival direction of the primary particle and to obtain information
about its nature.

The idea of observing EECRs from space was put forward by Benson and
Linsley more than thirty years ago~\cite{Benson-Linsley-1981}.\footnote{One can
find earlier dates in the literature but we failed to find a published work
prior to this one.} They suggested to look earth-ward from a satellite flying on
a circular equatorial orbit at height 500--600~km and equipped with a
mirror 36~m in diameter with~$10'$ resolution and $\sim5000$ photomultiplier tubes
5~cm in diameter located at the focal surface of the mirror. They estimated that
the instrument will have a circular field of view about 100~km in diameter (thus
covering an area three times larger than that of Auger), a duty cycle of the
order of 20--30\%, and the energy threshold below 10~EeV.

While the basic idea might look simple, it was realized soon that such an
experiment poses a whole number of challenging problems both in technology and
science since one has to register a faint flux of photons emitted in the
continuously varying conditions of the Earth atmosphere and background spoiled
by man-made lights, star light and other phenomena with an instrument working in
a harsh space environment and built with strong restrictions on mass, energy
consumption, etc.  Since the pioneering work by Benson and Linsley, different
aspects of registering EECRs from space have been studied in great detail, see,
e.g.,~\cite{modeling-2001,Pallavicini_etal-2012} and references therein.  Still,
not a single project of an orbital detector of EECRs has been implemented yet.
In what follows, we will review the current status of development of the TUS and
KLYPVE experiments and compare them with the most advanced of the existing
projects, JEM-EUSO.

\section{TUS} 

The TUS (Tracking Ultraviolet Setup) experiment was started in 2001~\cite{TUS1}
as a pathfinder for a more advanced KLYPVE project.
Plans of accommodating the TUS detector on a satellite were changed several
times and the detector itself was improved in comparison with the initial
version~\cite{TUS2, TUS3, TUS4}. 
Now it is accommodated on the Lomonosov satellite as a part of instrumentation
for studying the Extreme Universe Phenomena~\cite{TUS5} and is scheduled for
launching in late 2015.

\begin{figure}[!ht]
	\centerline{\includegraphics[width=.6\textwidth]{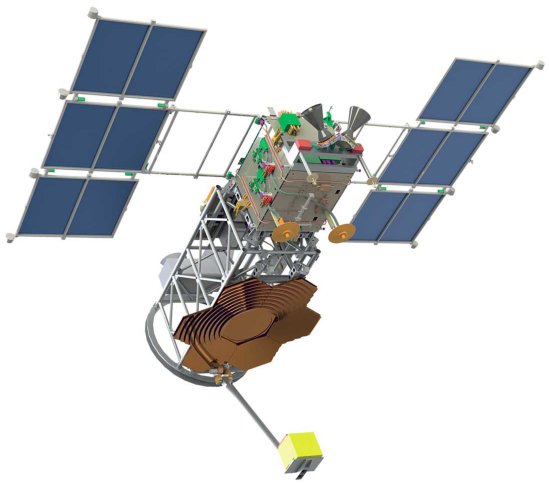}}
	\caption{Detector TUS on board the Lomonosov satellite.}
	\label{fig1:tus}
\end{figure}

The TUS detector is based on the same optical scheme as the one suggested by
Benson and Linsley but with a much more modest size of the elements.
It consists of two main parts: a mirror-concentrator with an area of
2~m$^2$ and a photodetector composed of 256 pixels, located at the mirror
focus, see figure~\ref{fig1:tus}.
The technology of TUS was described elsewhere~\cite{TUS6, TUS7, TUS8, TUS9} and 
only the main instrumental parameters are discussed below.

\begin{figure}[!ht]
	\centerline{\includegraphics[width=.5\textwidth]{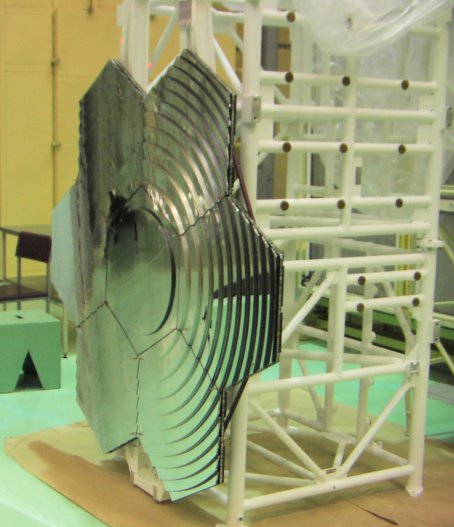}}
	\caption{Mirror-concentrator of the TUS detector attached to the Lomonosov frame.}
	\label{mirror}
\end{figure}

The mirror-concentrator is designed as an ensemble of 11 parabolic rings
that focus a parallel beam to one focal point. In this design, the thickness of the
mirror construction is small (3~cm) which is important for its deployment
on the satellite frame.  The focal distance of the mirror is 1.5~m.
The mirror is cut into hexagonal segments with diagonal 63~cm. The segments are
made of carbon plastic supported by a honey comb aluminum plate so that the
whole construction is stable in a wide range of temperatures. The surface of the
segments is obtained as plastic replicas of aluminum press forms (one for the
central mirror part and one for 6 lateral parts). In a vacuum evaporation
process, the plastic mirror surface is covered by aluminum film and protected by
a MgF$_2$ coat.  Reflectivity of the mirror surface at wavelength 350~nm (average
for the atmosphere fluorescence) is 85\%. The mirror passed various space
qualification as well as optical tests. Those tests demonstrated sufficient
stability of its optical quality in space conditions.
Expected life time of the mirror is more than 3 years.
The main parameters of the detector are listed in table~\ref{Table1}.
The mirror placed on the scientific payload frame of the Lomonosov satellite is
shown in figure~\ref{mirror}.

\begin{figure}[!ht]
	\centerline{\includegraphics[width=\textwidth]{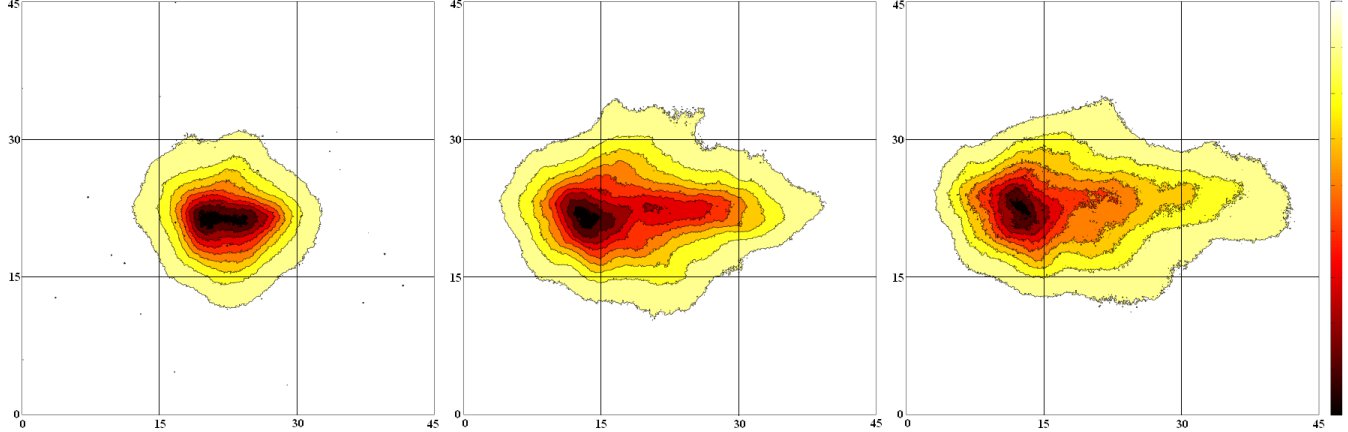}}
	\caption{Results of the PSF measurement. Zenith angles from left to right:
		$0^\circ$, $3^\circ$,
	$4.5^\circ$. Dimensions are in millimeters. A square of $15\times15$~mm$^2$
	corresponds to one pixel of the TUS photodetector.}
	\label{PSF}
\end{figure}

Optical tests were provided by two
methods: 1)~independent measurements of each mirror segment by scanning with
parallel laser beams (described in detail in~\cite{NIMa2014}, 2)~measurements
with a distant ($\sim$30~m) pinpoint light source. The first method allows one
to evaluate the quality of production of the mirror segments surface.  By the
second method, final results on the point spread function (PSF) of the
full size mirror at various beam angles were
obtained. They are shown in figure~\ref{PSF}.

\begin{table}
\caption{\label{Table1}Parameters of the TUS detector}
\begin{center}
\begin{tabular}{ll}
\br
Parameter, units&Value\\
\mr
Area of the mirror-concentrator, m$^2$&2\\
Mirror focal distance, cm&150\\
Pixel size, mm&15\\
Pixel number&256\\
Time step, $\mu$s&0.8\\
Observable area of the atmosphere at orbit  height 500~km, km$^2$&6400\\
Pixel size in the atmosphere, km&5\\
\br
\end{tabular}
\end{center}
\end{table}

The photodetector pixels are made by photomultiplier tubes (PMTs) R1463 of
Hamamatsu with multi-alcali cathode of 13~mm diameter. The quantum efficiency of the
PMT cathode is 20\% for wavelength 350~nm.  The PMT’s multi-alcali cathode (instead
of usually used bi-alcali one in ground-based fluorescence detectors) was chosen
because the cathode operates in a wider range of temperatures in the linear
regime. Light guides with a square entrance ($15\times15$~mm) and circle output
adjusted to PMT cathodes were used for making the detector field  of view
uniformly filled with pixels.

\begin{figure}[!ht]
	\centerline{\includegraphics[width=.65\textwidth]{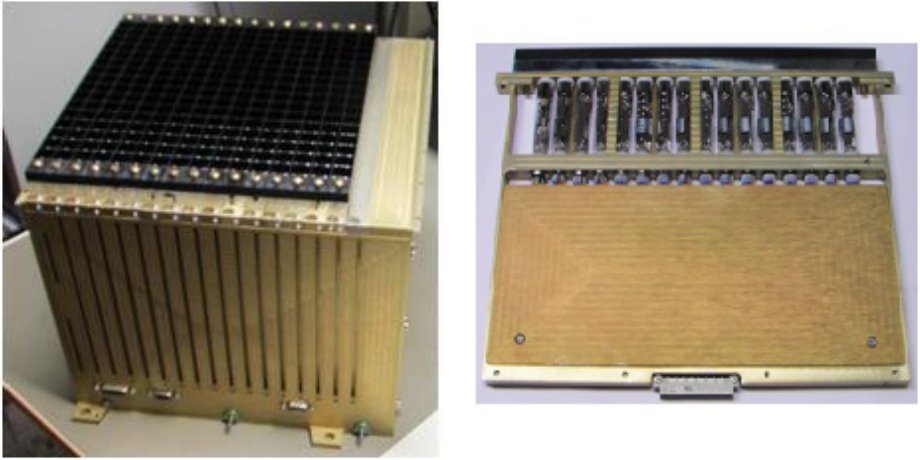}}
	\caption{The TUS photodetector (left) and one PMT cluster (right).}
	\label{PDM}
\end{figure}

After preliminary testing, PMTs with a similar gain were grouped in 16 clusters.
Data from each tube in a cluster are digitized by an analog-digital converter
and then analyzed and memorized by a field-programmable gate array
(FPGA).  The final detector triggering
and memorizing of all data is done by the central FPGA. Information volume of
one EAS data is $\sim$100~Kbytes. Expected volume of one day EAS data
transmitted to the mission center is 250~Mbytes.
It is worth noticing that electronics of TUS is developed to measure not
only EAS but also other transient phenomena in the atmosphere (lightnings,
transient luminous events, meteors). This is achieved by recording waveforms
of events with various time of digital integration (from 0.8~$\mu$s up to
6.6~ms). The duration of measured events varies from 205~$\mu$s up to
1.7~s.
The TUS photodetector and one PMT cluster can be seen in figure~\ref{PDM}.

Performance of the TUS detector was simulated taking into account parameters of
the real TUS mirror-concentrator and TUS electronics~\cite{TUS10}. The EECR 
trigger operates in two levels: at the first level, pixels with signals~$Z$ times
larger than noise RMS are selected. 
At the second level, signals triggered on the first level and lined up in space
in~$T$ consecutive time intervals are selected.
Data from all pixels are recorded by a command of the second level trigger. 
Numbers~$Z$ and~$T$ will be set from the mission center for the optimal
relation between the lowest TUS detector energy threshold and the trigger
rate, limited by volume of information to be transmitted to the mission center.

\begin{figure}[!ht]
	\centerline{\includegraphics[width=.6\textwidth]{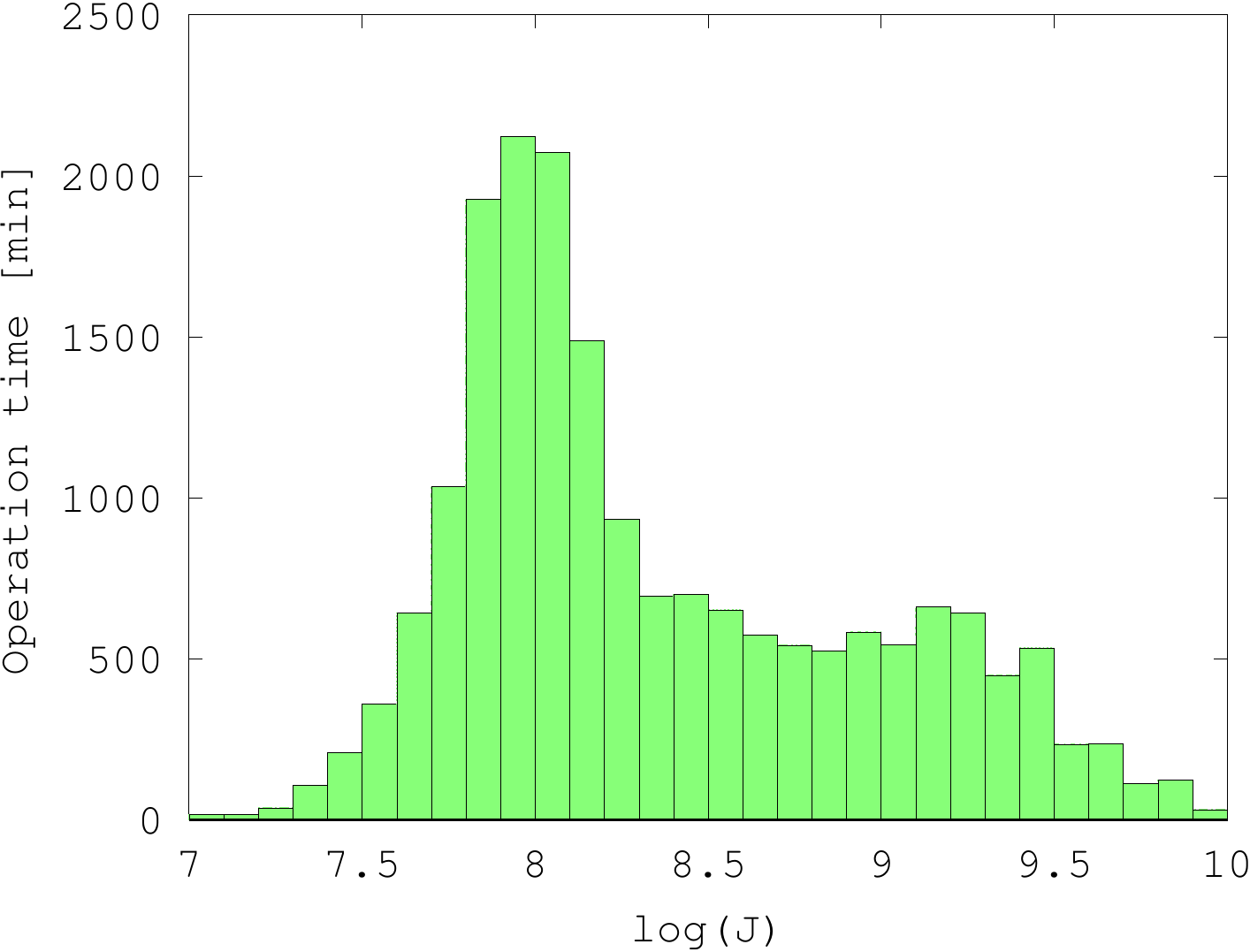}}
	\caption{Operation time of Tatiana-2 as a function of the measured
		atmospheric background intensity~$J$.}
	\label{fig:T2}
\end{figure}

Experimental results on the atmosphere glow were obtained during the
Universitetsky-Tatiana-2 mission (Tatiana-2)~\cite{TUS11}. 
The glow varies with the Moon phase and its height above the horizon but the atmosphere
emits a glow even at moonless nights. 
Data of the Tatiana-2 mission, which include approximately $1.9\cdot10^4$ minutes
of operation in night time, are presented in figure~\ref{fig:T2} as the
operation time versus measured UV background intensity~$J$.
The background intensity
varies from lower values at moonless nights (at the darkest atmosphere regions:
above the Pacific ocean, above deserts and a part of Siberia) with
$J=3\cdot10^7$--$10^8$~ph~cm$^{-2}$~s$^{-1}$~sr$^{-1}$ to full-moon nights with
$J\sim3\cdot10^9$~ph~cm$^{-2}$~s$^{-1}$~sr$^{-1}$.

The Lomonosov satellite will be launched to a polar orbit close to that of
Tatiana-2, so the available data can be used for estimating the TUS operation
time with various background intensities.  Results of such an estimate are
presented in table~\ref{Table2}.  The number of registered EECRs will be much
lower than those given in the table because the efficiency of the triggering
system was not taken into account.  With the limited exposure, the TUS detector
will not make a breakthrough to the problem of EECR origin. Its main aim is to
check the EAS fluorescence detector performance in a harsh space environment.

\begin{table}[!ht]
\caption{\label{Table2}Expected TUS performance parameters.}
\begin{center}
\begin{tabular}{llll}
\br
Intensity $J$,~ph~cm$^{-2}$~s$^{-1}$~sr$^{-1}$ & $3\cdot10^7$--$10^8$ & $10^8$--$5\cdot10^8$ & 
$5\cdot10^8$--$5\cdot10^9$\\
\mr
One year night time (2047 hrs)~vs~$J$, \% & 33 & 39 & 28\\
TUS one year exposure, km$^2$~year~sr &$3\cdot10^3$ & $3.7\cdot10^3$ & $2.7\cdot10^3$\\
Energy threshold (preliminary),~EeV & 60 & 150 & 400\\
EECR rate, events/km$^2$~year~sr & 0.024 & 0.002 & 0.0001\\
EECR event number per year & $\sim70$ &$\sim7$ & $<1$\\
\br
\end{tabular}
\end{center}
\end{table}

An important source of background events in orbital EECR measurements are UV
flashes (duration of 1--100~ms), whose origins are related to electrical
discharges in the atmosphere. In this respect, the data on UV flashes from the
Tatiana-2 satellite were analyzed~\cite{TUS12}. Measurements
were done in a  wide range of photon number~$Q$ in the atmospheric UV flashes: from
$Q=10^{21}$ up to $Q\sim10^{25}$ where tens of events were registered. The main
features of flashes with $Q>10^{23}$, i.e., their duration of 10--100~ms and their
global distribution concentrated in equatorial region above continents, allowed
suggesting that those flashes are either lightnings or transient luminous
events generically related to lightnings. Those ``bright'' flashes will be
easily separated from EAS fluorescent signals due to their long duration and an
enormous number of photons (to compare with EAS parameters: their duration is
not more than 0.1~ms and the number of UV photons $Q\sim10^{16}$ for $E=100$~EeV). 

More dangerous for imitation of EECR events is the background of dim and short
flashes ($Q\sim10^{21}-10^{23}$, duration $\sim$1~ms) also observed by the
Tatiana-2 detector. 
In contrast to bright flashes, dim ones are distributed more uniformly over the
Earth. Rates of both kinds of flashes (about 10$^{-4}$--$10^{-3}$~km$^{-2}$~hr$^{-1}$)
are much higher than the expected rate of EECR events
$\sim10^{-6}$~km$^{-2}$~hr$^{-1}$,
so that the problem of distinguishing EAS flashes from
atmospheric flashes is a complicated one. In this respect, the performance of the
TUS detector will be an important test of future space measurements of EECRs.

\section{KLYPVE}

\begin{figure}
	\begin{center}
		\includegraphics[width=.7\textwidth]{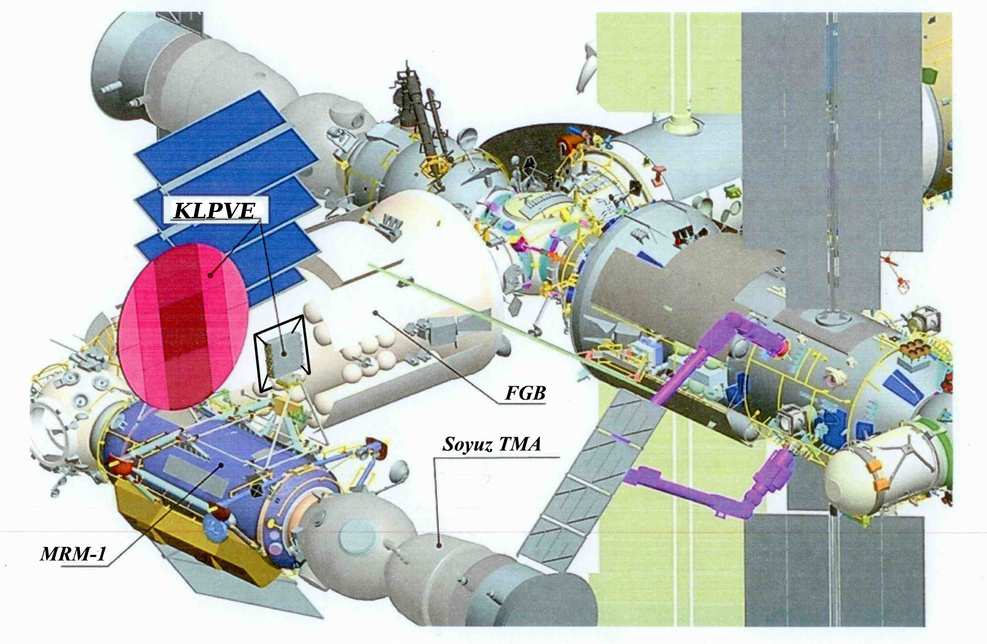}
	\end{center}
	\caption{\label{fig:KLYPVE_MRM}The KLYPVE detector on the MRM-1 module of the ISS.}
\end{figure}

In 2012, Skobeltsyn Institute of Nuclear Physics of Moscow State University
finished the preliminary design stage of the KLYPVE\footnote{KLYPVE is a Russian
acronym for extreme energy cosmic rays.} reflector type telescope for EECR
measurements from the ISS.  It will be located on the outer side of the Russian
segment, see figure~\ref{fig:KLYPVE_MRM}. The main  component of the detector is
a segmented optical system (OS) with a large entrance pupil. All parts of the
telescope are to be delivered to the ISS by the Progress-TM vehicle. 

\begin{figure}
	\begin{center}
		\includegraphics[height=7cm]{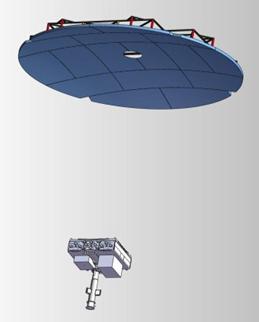}\quad
		\includegraphics[height=7cm]{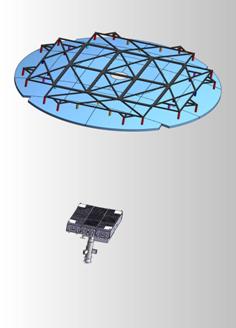}
	\end{center}
	\caption{\label{fig:KLYPVE}KLYPVE segmented mirror-concentrator and
		photodetector.  Left: view from the front side, right: view from the rear side.
		A supporting frame is shown at the rear side of the mirror.}
\end{figure}

The main goal of KLYPVE is to considerably increase the field of view (FOV) of
the detector in comparison with TUS and to decrease the energy threshold.  The
mirror-concentrator has a reflective surface of about 10~m$^2$ (entrance pupil
diameter 3.6~m, see figure~\ref{fig:KLYPVE}).  The OS has a focal distance of
3~m and FOV diameter of~15$^\circ$.  The angular resolution of the OS is the
same as in the TUS detector, i.e., 5~mrad (pixel size
$15\,\mathrm{mm}\times15\,\mathrm{mm}$).

However, it became clear during the preliminary design phase that
the characteristics of the instrument (observation area and image quality) do
not allow solving the problems related to anisotropy of EECRs.
The main reason for that is that due to the off-axis aberration of a fast
optical system, the spot size increases very rapidly with increasing field angle.
For an accurate reconstruction of EECR parameters (primary energy and
direction), it is desirable to make a pixel of the photodetector smaller. 
Thus one has to look for optics with a much smaller spot size.
These considerations initiated the development of a new OS for the KLYPVE
detector in order to increase the FOV and to improve the spatial and angular
resolution and the overall performance of the instrument.
This work began in a close collaboration with members of the JEM-EUSO
Collaboration in late 2013.

To eliminate the off-axis aberration, an additional corrective element was
introduced into the telescope system in form of a Fresnel lens.  The thickness
of the lens should be sufficiently small, 1~cm in this case.  The material of
the lens is PMMA-000 (by Mitsubishi Rayon Co., Ltd.), which is a UV-transparent
version of polymethyl metacrylate.  It has a good UV transmittance (more than
90\% for 15~mm thickness layer at wavelength more than 320~nm), with a
refractive index of about~1.51 and has been used in space on many occasions. The
lens has a radial Fresnel structure with groove depth 1~mm. Lenses with similar
characteristics were manufactured at Ohmori Materials Fabrication Laboratory,
RIKEN (Japan) in 2009--2011~\cite{JEM-lens-2011}.

Depending on the size and complexity of the forms of individual optical elements
of the system, two approaches have been proposed.

\subsection{Baseline System}

\begin{figure}
	\begin{center}
		\includegraphics[width=.52\textwidth]{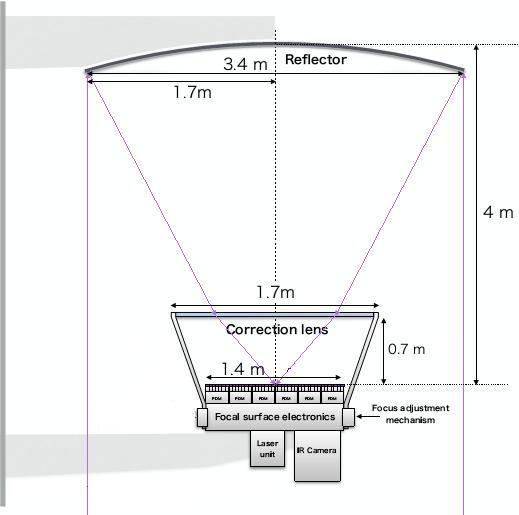}
		\includegraphics[width=.45\textwidth]{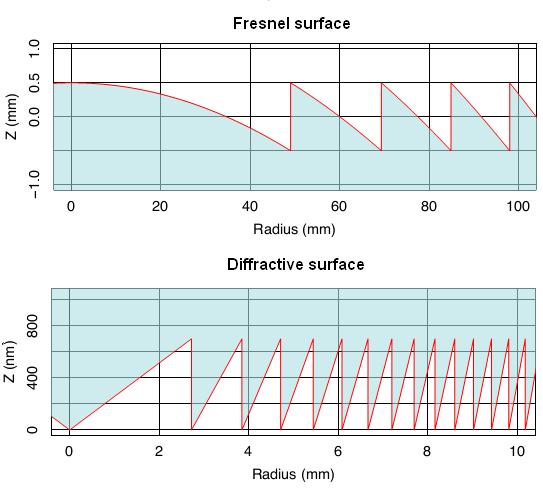}
	\end{center}
	\caption{\label{fig:BaseLine}The optics of the Baseline system.
		Right: grooves of the Fresnel and diffractive structures of the corrective lens.}
\end{figure}

In the so called Baseline system, the diameter of the reflector and the
lens-corrector equals 3.4~m and 1.7~m respectively. The total length of the
system (more precisely, the axial distance from the pole to the center of the
focal surface) is equal to 4~m, the distance from the lens to the focal surface
equals 70~cm, see figure~\ref{fig:BaseLine}. In this case, it is possible to
expand the FOV up to~$\pm 14^\circ$, and the diameter of the image is not larger
than 6~mm in the entire FOV.  The angular resolution of the system is
$\approx0.057^\circ$, which is equivalent to $\sim0.4$~km at ground.

One disadvantage of this approach is the difficulty of manufacturing of optical
elements, especially lenses. To correct aberrations, both lens surfaces should
have a small curvature, wherein the front side is a Fresnel spherical surface
(with a radial structure of grooves of 1~mm depth), and the back side is a
spherical diffraction surface (with grooves depth 700~nm, see the right side of
figure~\ref{fig:BaseLine}). 

Two main optical performance characteristics are the spot size and optical
efficiency, which can be determined by non-sequential ray tracing.
Table~\ref{tab:BL_performance} represents these parameters for different field
angles as RMS diameter of the focal surface image and as the ratio of the number
of rays in the spot to the number of rays at the entrance pupil.

\begin{table}
\caption{\label{tab:BL_performance}
Baseline OS performance characteristics for different field angles.}
\begin{center}
\begin{tabular}{lllllllll}
\br
Field Angle & 0$^\circ$ & 4$^\circ$ & 6$^\circ$ & 8$^\circ$ & 10$^\circ$ & 12$^\circ$ & 13$^\circ$ & 14$^\circ$ \\
\mr
Spot Size, mm & 2.8 & 2.6 & 3.2 & 4.2 & 4.5 & 4.3 & 4.0 & 3.9 \\
Optical Efficiency, \% & 62.8 & 61.7 & 58.8 & 57.4 & 51.4 & 43.8 & 37.1 & 29.6\\
\br
\end{tabular}
\end{center}
\end{table}

Delivering such a huge system to the ISS is a complicated task by itself.  In
case Progress-TM is used, cargo must be first placed inside the ISS and thus
pieces of the instrument have to pass through a cylindrical lock of 70~cm
diameter and 120~cm length.  The currently offered solutions require
segmentation of all the major components of the system, including the lens, the
mirror and the photodetector, and subsequent deployment in space.

\subsection{Multi-Eye Telescope System}

The optical system of the KLYPVE is a wide-field large aperture and fast optics
system.  For such a complex instrument, a more promising option might be a
Multi-Eye Telescope System (METS).  The main idea of METS is to divide a wide
FOV into several FOVs of identical smaller telescopes. In this case, we have a
simplification of an individual telescope: the design is simplified by cutting
the aperture and the FOV of an individual telescope compared to the Baseline
option thus reducing the requirements for their manufacturing. 

This system has several additional advantages: 
\begin{itemize}

	\item smaller dimensions of individual telescopes and their narrower FOVs
		allow correcting aberrations without the use of a complex surfaces corrector
		(diffraction, curved Fresnel);

	\item not only manufacturing of the mirror segments and lenses but testing and
		adjustment of the overall system might be simpler;

	\item dimensions of the elements can be chosen basing on the possibility of
		transporting the individual telescopes assembled.

\end{itemize}
One of the possible drawbacks of METS is the necessity of a trigger working for
all three telescopes simultaneously.

\begin{figure}
	\begin{center}
		\includegraphics[height=7.8cm]{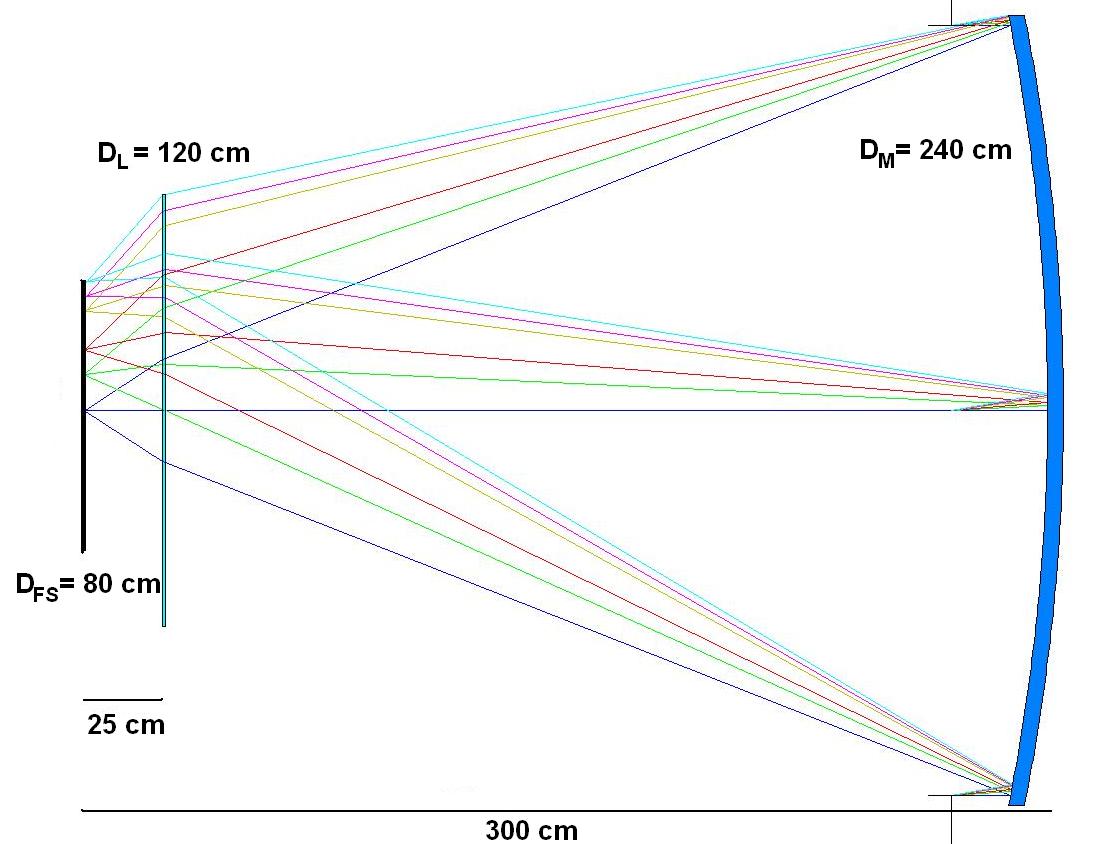}\qquad
		\includegraphics[height=6.cm]{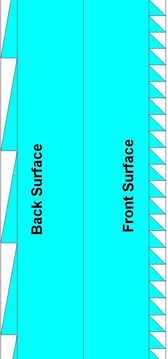}
	\end{center}
	\caption{\label{fig:METS}Left: the METS optical scheme. 
		Right: grooves of the Fresnel structures at the edge of the corrective lens.}
\end{figure}

In the optimization calculations, the following dimensions of an individual
telescope were obtained: 2.4~m diameter mirror, 1.2~m lens, 0.9~m photodetector,
the total axial length 3~m, see figure~\ref{fig:METS}.  Correction of
aberrations in the FOV $\pm 10^\circ$ (which corresponds to the diameter of
overall FOV of three telescopes $\sim35^\circ$) can be achieved using plane
Fresnel surfaces.  The angular resolution of METS is $\approx0.075^\circ$, which
is equivalent to $\sim0.5$~km at ground.

\begin{figure}
	\hspace{25mm}
	\includegraphics[width=7.4cm]{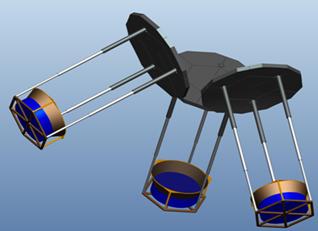}\quad
	\includegraphics[width=3.2cm]{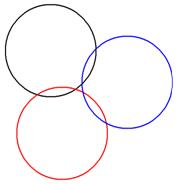}

	\phantom{}\hspace{25mm}
	\includegraphics[width=7.4cm]{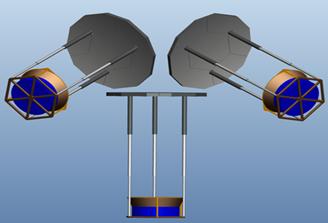}\quad
	\includegraphics[width=4.5cm]{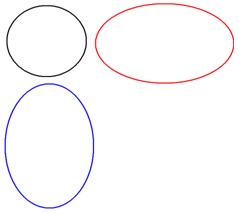}
	\caption{\label{fig:METS_modes}Top: the basic mode of METS operation. The angles
		of inclination of individual telescopes are exaggerated for clarity. 
		Bottom: The tilted mode of METS operation.
		The right panels show mutual position of FOVs of separate telescopes.}
\end{figure}

Spot size and optical efficiency of the METS telescope is presented in
table~\ref{tab:METS_performance}.  The sharp increase of the spot size on the
edge of the FOV is due to the fact that a flat focal surface was used
in the calculations.

\begin{table}[!ht]
	\caption{\label{tab:METS_performance}The METS OS performance characteristics for
	different field angles.}
\begin{center}
\begin{tabular}{lllllllll}
\br
Field Angle & 0$^\circ$ & 3$^\circ$ & 5$^\circ$ & 6$^\circ$ & 7$^\circ$ & 8$^\circ$ & 9$^\circ$ & 10$^\circ$\\
\mr
Spot Size, mm & 2.0 & 2.7 & 3.7 & 3.7 & 4.0 & 4.2 & 4.5 & 5.4 \\
Optical Efficiency, \% & 68.8 & 67.2 & 62.0 & 57.7 & 55.0 & 53.3 & 49.1 & 41.9\\
\br
\end{tabular}
\end{center}
\end{table}

To control spots within 5~mm, one should slightly bend the focal surface at the
edges.  Another advantage of the multi-eye system is the ability to use an
active (dynamic) configuration.  A special support structure of three telescopes
can afford one to implement operation of the detector in various modes. In the
basic mode, shown in the top row of figure~\ref{fig:METS_modes}, the FOVs of
individual telescopes are adjacent (or slightly overlapping for
cross-calibration). In the coincidence mode (overlapping FOVs), one can
significantly reduce the energy threshold. Finally, in the tilted mode (FOVs of
two or all telescopes are inclined relative to the direction at the nadir, see
the bottom row of figure~\ref{fig:METS_modes}), it is possible to increase the
exposure, and hence to collect larger statistics of the EECR events.

\section{JEM-EUSO}

\begin{figure}[!ht]
	\centerline{\includegraphics[width=.7\textwidth]{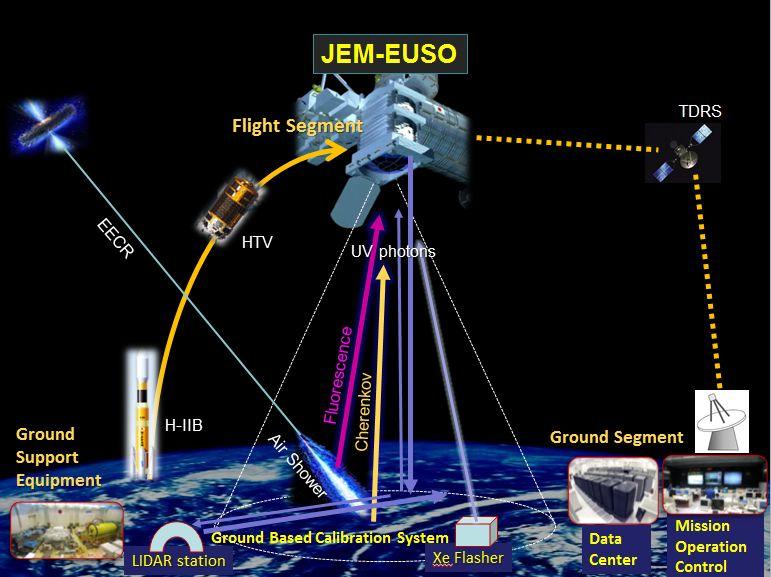}}
		\caption{Principle of detecting EECRs with the JEM-EUSO
	telescope~\cite{JEM-EUSO-2014}.}
	\label{fig:jem}
\end{figure}

JEM-EUSO (the Extreme Universe Space Observatory on board the Japanese
Experiment Module), which aims to be installed at the International Space
Station (ISS), is the most ambitious and sophisticated of the three projects,
see figure~\ref{fig:jem}.

The instrument consists of a UV telescope and an atmospheric monitoring system.
The telescope consists of three Fresnel lenses with an optical aperture of
4.5~m$^2$ and a focal surface detector formed by 137 photodetector modules
composed of $\sim5000$ multi-anode photo-multiplier tubes in total. The focal
surface detector thus includes $\sim3\cdot10^5$ channels providing a spatial
resolution $\approx0.074^\circ$, equivalent to $\sim0.5$~km at ground.  The
opening angle of the telescope equals $60^\circ$ providing an observational area
of $\sim1.4\cdot10^5$~km$^2$ in nadir mode.  The annual exposure of JEM-EUSO for
EECRs above 100~EeV is estimated to be an order of magnitude larger than that of
Auger~\cite{JEM-exposure-2013}.  Thanks to the ISS orbit, JEM-EUSO is able to
survey the entire celestial sphere almost uniformly, with the level of
non-uniformity of the exposure on declination and right ascension
$\sim10$\%~\cite{JEM-exposure-2014}.  Detailed simulations show that due to
these factors JEM-EUSO will be able to detect significant anisotropies of EECRs
practically in all feasible astrophysical
scenarios~\cite{Parizot_etal-anisoexpect-2014} thus providing a huge step
towards finding sources of EECRs.  In December 2013, the Japanese Space Agency
(JAXA) refused to deploy the instrument on board the JEM.  A detailed discussion
of the current status of the project and its pathfinders can be found
in~\cite{Bertaina-Parizot-2014} and in dedicated contributions by A.~Haungs and
M.~Fukushima in this volume.

\section{Discussion}

\begin{figure}[!ht]
	\begin{center}
		\includegraphics[width=12cm]{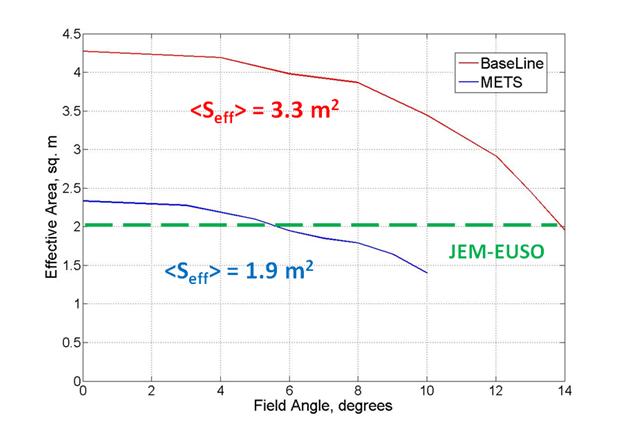}
	\end{center}
	\vspace{-5mm}
	\caption{\label{fig:EffArea}Effective area of the Baseline system
		and METS compared to JEM-EUSO for the entire FOV.
		$\langle S_\mathrm{eff}\rangle$ denotes
		average values of the effective area for the isotropic distribution of light.}
\end{figure}

Figure~\ref{fig:EffArea} shows the dependence of the effective
area~$S_\mathrm{eff}$ of the Baseline and METS optical systems of the modified
KLYPVE detector and the JEM-EUSO telescope on the field angle~$\gamma$.  Here,
the effective area means the ratio of the luminous flux to irradiance, i.e.,
$$
S_\mathrm{eff} = \varepsilon_\mathrm{opt}(\gamma) S_\mathrm{geom}(0^\circ)\cos\gamma,
$$
where $\gamma$ is the field angle, $\varepsilon_\mathrm{opt}$ is the optical efficiency,
and the axial geometrical area (the second factor) takes into account the
reduction of the entrance pupil due to the central screening by the lens. Thus,
this parameter is one of the most important in the design of optical systems of
an orbital detector.

The effective area is reduced at the edge of FOV for both systems, primarily
due to an increase in the scattering on Fresnel and diffraction (in case of the
Baseline system)
grooves, which leads to a significant spot smearing at large field angles. As can
be seen from figure~\ref{fig:EffArea}, the effective area of the METS telescope
approximately coincides with the JEM-EUSO one.
In the case of the Baseline system,
it is possible to increase the value of this parameter in average over the
entire FOV by 70\%, and more than twice for events in the cone $\gamma <10^\circ$.
However, it should be noted that a refraction efficiency of the (first)
diffraction peak was assumed to be 95\% in these calculations.
Reaching such values over the entire surface during manufacturing of the lens
can pose a problem since the diffractive surface area is about 2.3~m$^2$.
Thus, having undoubted advantages in decreasing the energy threshold, the
Baseline system might have the
effective FOV much smaller than the total FOV of the METS telescope.

\begin{table}
	\caption{\label{tab:OptPerformance}Performance characteristics (spot size~$d$,
		effective area~$S_\mathrm{eff}$, FOV~$\Omega$ and throughput
		$\Omega S_\mathrm{eff}$) for 5 different orbital detectors of EECRs.}
\begin{center}
\begin{tabular}{lllll}
\br
 & $d$, mm & $S_\mathrm{eff}$, m$^2$ & $\Omega$, sr & $\Omega S_\mathrm{eff}$, m$^2$\,sr \\
\mr
JEM-EUSO & 3 & 2 & 0.8 & 1.6 \\
TUS & 20 & 1.2 & 0.025 & 0.03 \\
KLYPVE (orig.) &  20 & 5 & 0.05 & 0.25 \\
Baseline & 3.5 & 3.5 & 0.2 & 0.7 \\
METS & 4 & 2.0 & 0.3 & 0.6 \\
\br
\end{tabular}
\end{center}
\end{table}

\begin{figure}[!ht]
	\centerline{\includegraphics[width=.6\textwidth]{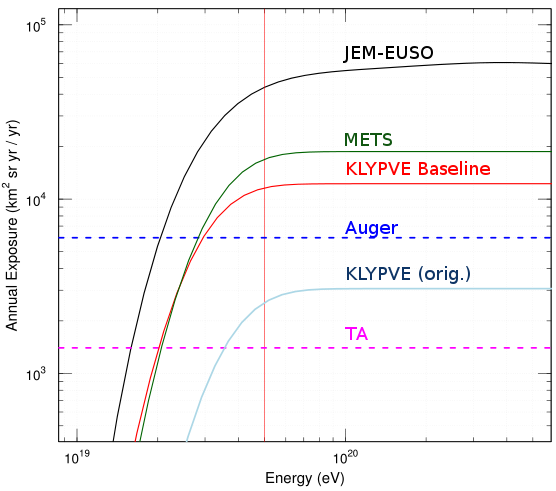}}
	\caption{Annual exposure of different ground-based and space experiments.} 
	\label{fig:exp}
\end{figure}

Table~\ref{tab:OptPerformance} summarizes the main characteristics of the five
optical systems: the lens system of JEM-EUSO, TUS Fresnel mirrors, the original
version of the KLYPVE detector (without any corrector) and the two advanced
options, the Baseline system and METS (with three identical telescopes in the
basic mode of operation). Scattering of light by Fresnel surfaces was taken into
account when calculating the effective area.  It is important to underline that
the effective area of the Baseline system provides an annual exposure two times
larger than that of Auger, see figure~\ref{fig:exp}.  Together with an almost
uniform exposure of both hemispheres, this gives an additional advantage for
anisotropy studies.

\section{Conclusions}

Three orbital projects aimed at detecting extreme energy cosmic rays are
being developed these days.\footnote{%
	A discussion of the Super-$\mathcal{EUSO}$~\cite{Super-EUSO-2011} and OWL~\cite{OWL-2013}
	missions goes beyond the scope of the article.}
TUS is scheduled for launching in late 2015. If
successful, TUS will become the pioneer of exploration of EECRs from space and a
pathfinder for more advanced missions.
It will also provide precious information about the UV background of the
Earth atmosphere.
The KLYPVE project finished the stage of
the preliminary design and is included in the Russian Federal Space Program. It
has entered the stage of the design project in its modified version aimed to
improve the technical parameters of the instrument.  There is a realistic possibility
for KLYPVE to be installed on board the Russian Segment of the ISS in the next
6--7 years.  The development of the full instrument of the JEM-EUSO project is currently facing
certain problems, but the successful flight of
EUSO-Balloon\footnote{See
\texttt{http://euso-balloon.lal.in2p3.fr}} on 24~August,
2014, which was used to test a number of components of the telescope and the atmospheric
monitoring system, gives a strong ground
for the hope that difficulties will be overcome and the most sophisticated of
the three projects will be implemented in the future.  Both KLYPVE and JEM-EUSO will
provide outstanding possibilities for a breakthrough in solving the long-standing puzzle of
the nature and origin of the most energetic particles ever detected on Earth.

\ack

We thank the members of the JEM-EUSO collaboration who are taking an
active part in the development of the modified KLYPVE project.
Our special thanks are addressed to Andreas~Haungs.
The work was done with a partial financial support by
the Russian Foundation for Basic Research grant 13-02-12175-ofi-m.
N.S. was supported by the ``Helmholtz Alliance for Astroparticle Physics HAP''
funded by the Initiative and Networking Fund of the Helmholtz Association,
Germany.

\section*{References}
\bibliographystyle{iopart-num}
\bibliography{ecrs}
\end{document}